\documentclass[12pt,preprint]{aastex}

\usepackage{natbib}
\usepackage{graphicx}
\shorttitle{M Dwarf Flares from SDSS Spectra}

\begin{document}

\title{M Dwarf Flares from Time-Resolved SDSS Spectra}

\author{Eric J. Hilton}
\affil{Astronomy Department, University of Washington,
    Box 351580, Seattle, WA 98195}
\email{hilton@astro.washington.edu}

\author{Andrew A. West}
\affil{Boston University, Department of Astronomy, 725 Commonwealth
  Ave, Boston, MA 02215}

\author{Suzanne L. Hawley and Adam F. Kowalski}
\affil{Astronomy Department, University of Washington,
    Box 351580, Seattle, WA 98195}

\begin{abstract}

We have identified 63 flares on M dwarfs from the individual component 
spectra in the Sloan Digital Sky Survey using a novel measurement
of emission line strength called the Flare Line Index.
Each of the $\sim$38,000 M dwarfs in the SDSS low mass star spectroscopic sample of West et al. was observed 
several times (usually 3-5) in exposures that were typically 9-25 minutes in duration.
Our criteria allowed us to identify flares that exhibit very strong 
H$\alpha$ and H$\beta$ emission line strength and/or significant 
variability in those lines
throughout the course of the exposures.
The flares we identified have characteristics consistent with flares
observed by classical spectroscopic monitoring.
The flare duty cycle for the objects in our sample is found to increase 
from 0.02\% for early M dwarfs to 3\% for late M dwarfs.
We find that the flare duty cycle is larger in the population near 
the Galactic plane and that the flare stars are more spatially restricted 
than the magnetically active but non-flaring stars.  
This suggests that flare frequency may be related to stellar age (younger stars are more likely to flare)
and that the flare stars are younger than the mean active 
population.
 
\end{abstract}


\section{Introduction}
\label{sec:intro}

Flares are explosive events caused by magnetic reconnection in stellar atmospheres \citep[and references therein] {Haisch1991}. 
In the standard model, electrons are accelerated along magnetic field
lines and impact the lower stellar atmosphere.
This catastrophic release of magnetic energy causes emission from the radio to the X-ray \citep{Hawley1995, Hawley2003, Osten2005, Fuhrmeister2007, Berger2008}.
On M dwarfs, the most distinct observational characteristic is the tremendous increase in the
blue and near-ultraviolet continuum emission, 
up to several magnitudes in a few minutes or less
\citep{Hawley1991,Eason1992, Kowalski2010}.
The initial burst, or impulsive phase, is also characterized in the optical spectrum by 
increases in chromospheric line emission, particularly in the 
hydrogen Balmer lines \citep{Hawley1991, Mart'in2001, Fuhrmeister2004}.
The impulsive phase is followed by a gradual decay phase which 
can last from tens of minutes to hours for the largest 
flares \citep{Moffett1974, Zhilyaev2007}. 

The rate at which flares occur is of significant current interest.
Recently, numerous planets have been discovered around M dwarfs, 
including some of only a few Earth masses \citep{Udry2007,Charbonneau2009, Forveille2009, Mayor2009, Correia2010}.
The atmospheres of these planets may be greatly affected by the amount of high-energy radiation incident upon them.
Cool, red M dwarfs produce little high-energy radiation during quiescence, so the flaring rate is an important factor in determining planet habitability
\citep{Heath1999, Segura2005, Tarter2007, Walkowicz2008, Segura2010}.

Additionally, characterizing the M dwarf flaring rate is important 
for new instruments such as Pan-STARRS \citep{Kaiser2004}, the Palomar
Transit Factory \citep{Rau2009} and the Large Synoptic Survey Telescope 
\citep{LSST2009}, which will carry out large, all-sky surveys in the 
time-domain.
Effectively selecting rare, exotic transients from the large sample of 
M dwarf flares that will be observed in these surveys
requires reliable knowledge of expected flare characteristics and rates.

Traditionally, the frequency of flares on M dwarfs has been determined through time-resolved photometric monitoring 
of individual stars.
\citet{Lacy1976} and \citet{Gershberg1983} have amassed several
hundred hours of optical observations on dozens of the most
magnetically active and well-known
flare stars and have derived power law relationships between the total energy in the \emph{U} band and the frequency of a flare, with less energetic flares occurring more frequently.
Studies on individual stars have confirmed the power law form of the frequency 
distribution, albeit with a range
of exponents \citep{Walker1981, Pettersen1984, Robinson1995, Leto1997, Robinson1999}.
At ultraviolet wavelengths,
\citet{Audard2000} and \citet{Sanz-Forcada2002} 
found that the EUVE flare frequency distributions for active stars also
had a power law form.  
In addition, \citet{Audard2000} demonstrated that
large flares occur preferentially on the X-ray brightest stars.

Spectroscopic studies of flares have also investigated the flare rate,
but with smaller samples and incomplete information on flare energies
it is not possible to compute the flare frequency distribution as
described above for
photometric data.  Instead
the results are usually presented as ``flare duty cycles'', meaning that
during the entire period of observation, flares occur during a certain percentage of the time.
Alternatively a ``flare rate'' (number of flares per hour) is sometimes given, but without reference
to the energy distribution of the flares.   
Both the flare duty cycle and flare rate are strong functions of the 
observational sample, since such factors as the individual exposure
time (compared to the typical flare timescale), the flare visibility
in the line radiation compared to the continuum, and the flare energy
will all be important in determining whether a particular observation
is counted as flaring.

\citet{Reid1999} estimated a duty cycle of 7\% from H$\alpha$ observations of the M9.5 dwarf BRI 0021-0214, while
\citet{Crespo-Chac'on2006} observed 14 small flares on AD Leonis, 
none of which produced measurable continuum increase, and determined 
a flare rate of $> 0.7 \mbox{ flares}/\mbox{hr}$.
\citet{Schmidt2007} investigated a sample of 81 M dwarfs, each observed a few times and
found a flare duty cycle of $\sim$5\%, based on H$\alpha$ variability.
Recently, \citet{Lee2010} reported a duty cycle of $\lesssim$ 5\% for H$\alpha$ flares (defined as a brightness increase of a factor of ten)
from time-resolved spectroscopic monitoring ($\sim$1 hour each) of 43 M dwarfs.

The small numbers of stars in most previous studies (both photometric and spectroscopic) did not allow examination of whether the flare rate is a 
function of stellar or Galactic parameters.
However, \citet{Kowalski2009} used the repeat observations from the SDSS Stripe 82 photometry ($\sim$60 epochs per object spread over 10 years) to identify flares and determine flare rates.
They identified 271 flares on more than 50,000 M dwarfs, a sample which is orders of magnitude larger than previous work and the
 first to contain significant numbers of both magnetically active and inactive stars.
The vast majority of flares (but not all) in the \citet{Kowalski2009} study occurred on active stars, and they found that 
the flare star spatial distributions reflect the active star distributions 
as a function of spectral subtype \citep{West2008}.

It should be clear that the flare rates, flare duty cycles, and flare frequency 
distributions described above are distinctly different measurements, and 
are difficult to compare.  In addition, the criteria used to define a flare
differ depending upon the method of observation (photometry vs spectroscopy),
the exposure time during which a flare may occur (resolved vs unresolved),
the wavelength (optical vs near-ultraviolet), and the emission mechanism (continuum
vs lines).  The sample selection (spectral type distribution, Galactic height distribution, color selection, etc.) also has a strong influence on the number of flares found.
Additionally, flares themselves are not homogeneous.
Flares with the same energy exhibit apparent magnitude variations that depend 
on the brightness of the star during quiescence
(the so-called ``contrast'' effect).
On stars with the same quiescent brightness, flares of the same energy 
exhibit a wide range of light curve shapes, with some flares rising and 
decaying quickly, giving large changes in apparent magnitude, 
while others are slower,
with smaller peak magnitude.  Understanding how often an 
M dwarf is seen in a flaring state requires knowledge of many stellar, 
flare, and sample properties.

Spectroscopic flare data typically have longer exposure times, may
contain unresolved flares, are usually at optical wavelengths, and are more likely
to observe flares, especially small ones, through enhanced line emission.
Flare rates and frequency distributions are therefore not good descriptions of
the data, and we will use the flare duty cycle (the fraction of epochs
classified as flares) in this work.

Here we present the first large, statistical study of spectroscopic flare duty cycles.
The low mass stars in this study are taken from the Sloan Digital Sky Survey Data Release 5 \citep[SDSS DR5;][] {Adelman-McCarthy2007}
 low-mass star spectroscopic sample \citep{West2008}, which consists of over 38,000 objects between spectral types M0 and L0.
 All SDSS spectra are the result of coadding several (usually 3-5) individual exposures, typically (but not always) observed on the same night.
The individual component exposure times were generally between 9 and 25 minutes.
The individual spectra were made public as part of the SDSS Data Release 6 \citep{Adelman-McCarthy2008}.
The repeated, shorter exposures are used to aid in cosmic ray rejection for the final combined SDSS spectrum.
However, these repeat spectra can be used to examine the time variability of spectral features.
We take advantage of the time sampling contained in the individual exposures to identify flares based on the strength and variability of spectral lines.

We have created a unique tool we use to identify flares: the Flare Line Index (FLI), 
which allows us to quantitatively measure spectral line strength in low signal-to-noise (SNR) spectra, where equivalent width measurements are not possible.
We then used False Discovery Rate analysis \citep{Miller2001} to identify flares based on exceptionally strong emission line strength.
Additionally, we used FLI variability criteria to identify flares based on changes in emission line strength over the course of the exposures.

In \S \ref{sec:sample} we further characterize our sample and describe our quality cuts.
The tools for our analysis are described in detail in \S \ref{sec:flare_tools}. The flare selection criteria are defined in \S \ref{sec:def_flare}, and
the results are presented in \S \ref{sec:results}.
Finally, we conclude in \S \ref{sec:conclusion}.

\section{The Sample}
\label{sec:sample}

We began with the SDSS DR5 low-mass star sample of \citet{West2008}, for which
low mass stellar candidates were chosen from the SDSS DR5 spectroscopic catalog using the photometric color selection $r - i > 0.53$ and $i - z > 0.3$,
 along with a 
standard set of SDSS processing flags (SATURATED, BRIGHT, NODEBLEND, INTERP\_CENTER, BAD\_COUNTS\_ERROR, PEAKCENTER, NOTCHECKED, and NOPROFILE were all required to be zero).
Stars with $r$-band extinction $>$ 0.5, colors consistent with having a white dwarf companion \citep{Smolv2004}, or measured radial
velocity $>$ 500 km s$^{-1}$ were culled from the sample (see \citet{West2008} for more details).

\begin{figure}
  \includegraphics[width=.8\textheight]{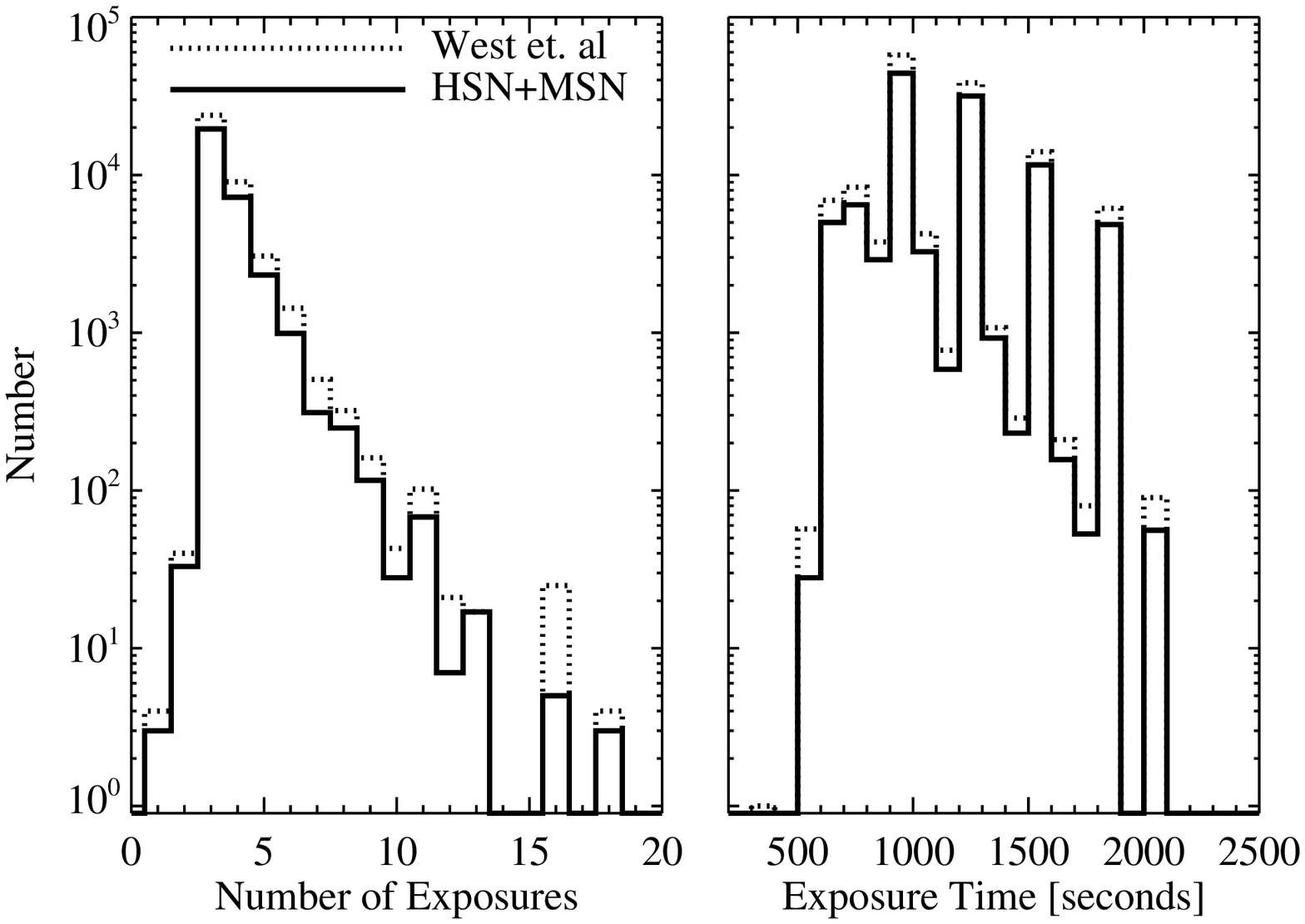}  
     \caption{Histograms of the exposure times for all component spectra in the sample (left) and of the exposure times for the component spectra (right).
     The dashed histograms are for the entire \citet{West2008} sample, while the solid histograms are the stars considered in this study.
     }
   \label{fig:exp_time_hist}
\end{figure}

The 38,717 spectra of individual objects in the DR5 sample that meet the above criteria consist of 142,032 individual exposures that were made available in DR6.
The distribution of number of exposures per object and the exposure times are shown in Figure \ref{fig:exp_time_hist}.
Although most of the spectra for an object were observed consecutively on a single night, 27\% of objects have additional exposures
separated by two hours or more.

\begin{figure}
  \includegraphics[width=.4\textheight]{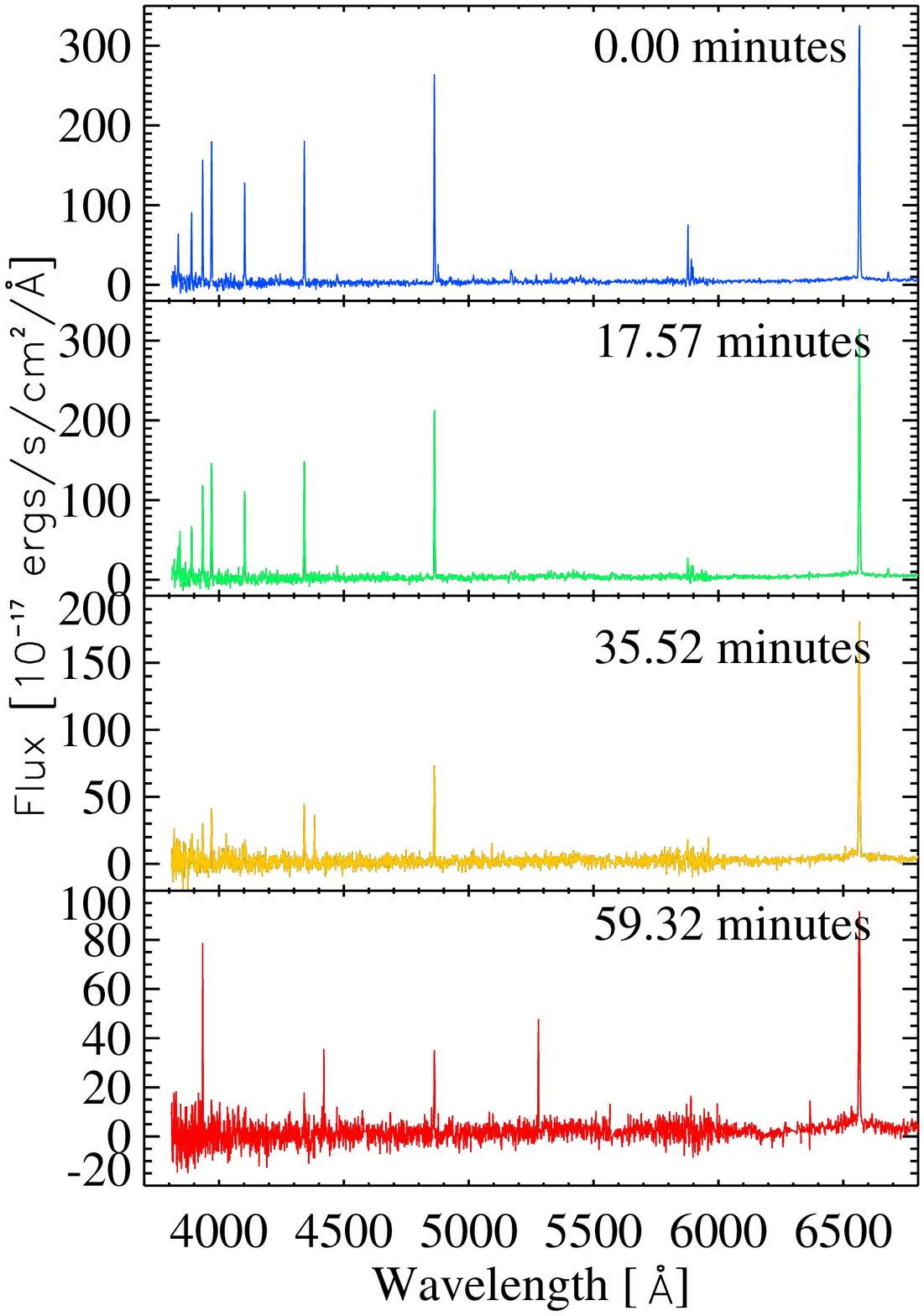}
  \includegraphics[width=.4\textheight]{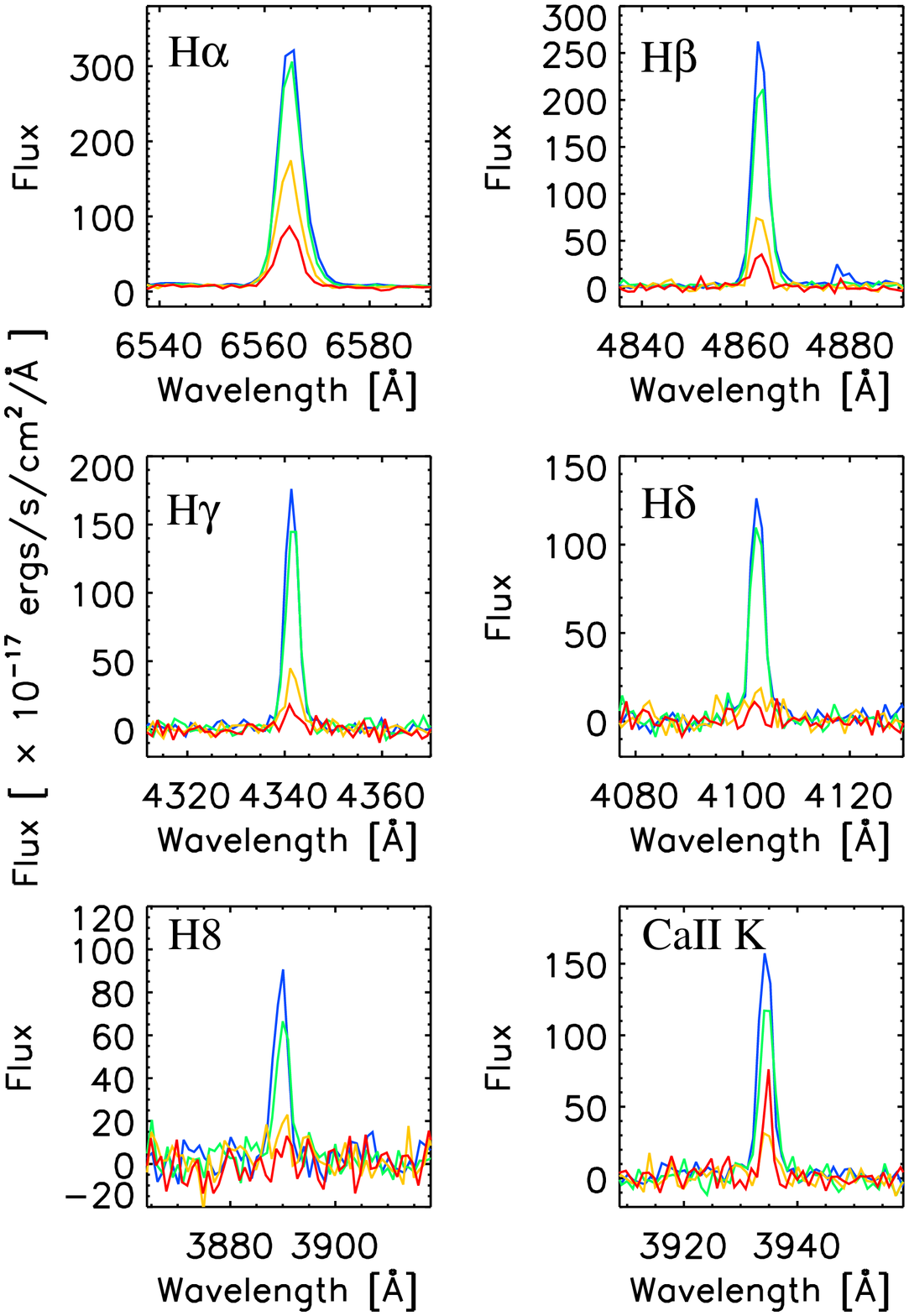}
  \caption{A sequence of spectra of a typical decay phase flare (left), with the time in minutes from the beginning of the first exposure.
  A detailed look at the emission lines (right, color corresponding to the exposure number on the left) reveals that H$\alpha$ and H$\beta$ 
  remain in emission longer than the higher-order Balmer lines.
  Note that Ca II K increases during the third exposure, an effect not observed in the Balmer lines.
        }
  \label{fig:four_spectra_and_lines}
\end{figure}

The SDSS spectra have wavelength coverage from 3800 to 9200 \AA, and resolution ranging from 1800 to 2200.
We therefore were able to measure five hydrogen Balmer lines (H8, H$\delta$,H$\gamma$, H$\beta$, H$\alpha$), as well as Ca II K.
Ca II H and H$\epsilon$ are blended at this resolution and are not considered
in our analysis.
The wavelength regions used for the continuum and line regions are given in Table 1 \citep[cf.][]{Bochanski2007}.
An example of the data, Figure \ref{fig:four_spectra_and_lines} shows the four exposures and line profiles from the 
six measured lines of a typical decay phase flare.

\begin{deluxetable}{llll}
\label{tab:wave_regions}
\tablewidth{0pt}
\tablecaption{Line and continuum regions}
\tablehead{
\colhead{Line} & \colhead{Line Region (\AA)} & \colhead{Continuum 1 (\AA)} & \colhead{Continuum 2 (\AA)} }
\startdata
H8 			& 	3884.15 - 3898.15  & 3850 - 3880  & 3910 - 3930 \\
Ca II K 		& 	3928.66 - 3938.66  & 3910 - 3925  & 3950 - 3960 \\
H$\delta$ 	&	4097.00 - 4110.00  & 4030 - 4080  & 4120 - 4170 \\
H$\gamma$ &	4331.69 - 4350.00  & 4270 - 4320  & 4360 - 4410 \\
H$\beta$	&	4855.72 - 4870.00  & 4810 - 4850  & 4880 - 4900 \\
H$\alpha$	&	6557.61 - 6571.61  & 6500 - 6550  & 6575 - 6625 \\
\enddata
\end{deluxetable}

\subsection{Quality Cuts}
\subsubsection{Cosmic Rays}
\label{sec:cosmic_rays}
Cosmic Rays (CRs) were not removed from the individual component spectra in the SDSS DR6 spectroscopic pipeline, but
individual pixels that contained likely CRs were flagged.
We visually inspected many of the flagged pixels and found that the pipeline routines frequently misidentified stellar emission lines as CRs. 
Since flares produce strong and time variable emission that might be misclassified as CRs, 
we developed a separate criterion to identify pixels as CRs in the line regions, so as not to reject genuine emission:
a pixel within the line region was called a CR if the difference between that pixel and the continuum level was more than ten times
the standard deviation of the other pixels in the region.
Our criterion identified CRs in the H$\alpha$ and H$\beta$ line regions in 0.55\% and 0.69\% of spectra respectively, 
as compared to 2.7\% and 3.5\% of the spectra that the SDSS pipeline flags identified.
There were no significant variations in the rate of CRs detected at different spectral subtypes using either definition.

Although this criterion still allows some CRs to remain undetected and masquerade as emission lines, 
we require emission in both H$\alpha$ and H$\beta$ to identify an object 
as in a flaring state (see \S \ref{sec:def_flare}).
This requirement greatly reduces the contamination by CRs, since 
two independent CR strikes on H lines in the same spectrum happen very
rarely.
Using the SDSS CR flags, we found that a ``CR'' is detected in both 
the H$\alpha$ and H$\beta$ regions in 0.3\% of the spectra, compared to
0.007\%  with our CR criterion.
All of the objects that met our flare criteria were inspected by eye, 
ensuring that no CRs led to an incorrect flare classification.
Since the continuum regions are much wider than the line regions and 
are only used as an indicator of the continuum level and signal-to-noise 
ratio (SNR),
we used the more conservative SDSS CR flags to mask any suspect pixels in all calculations involving the continuum regions.

\subsubsection{Sample Separation by Signal-to-Noise Ratio}
\label{snr_cuts}
The large sample size allowed us to divide our objects into 3 quality 
bins based on the SNR. Because M dwarfs are red and often have low 
SNR in the blue portions of their spectra,
our division into three bins identified those spectra that could be analyzed in both the H$\alpha$ and H$\beta$ regions.
These divisions were also important for our False Discovery Rate analysis (see \S \ref{sec:fdr}). 

\begin{figure}
  \includegraphics[width=.8\textheight]{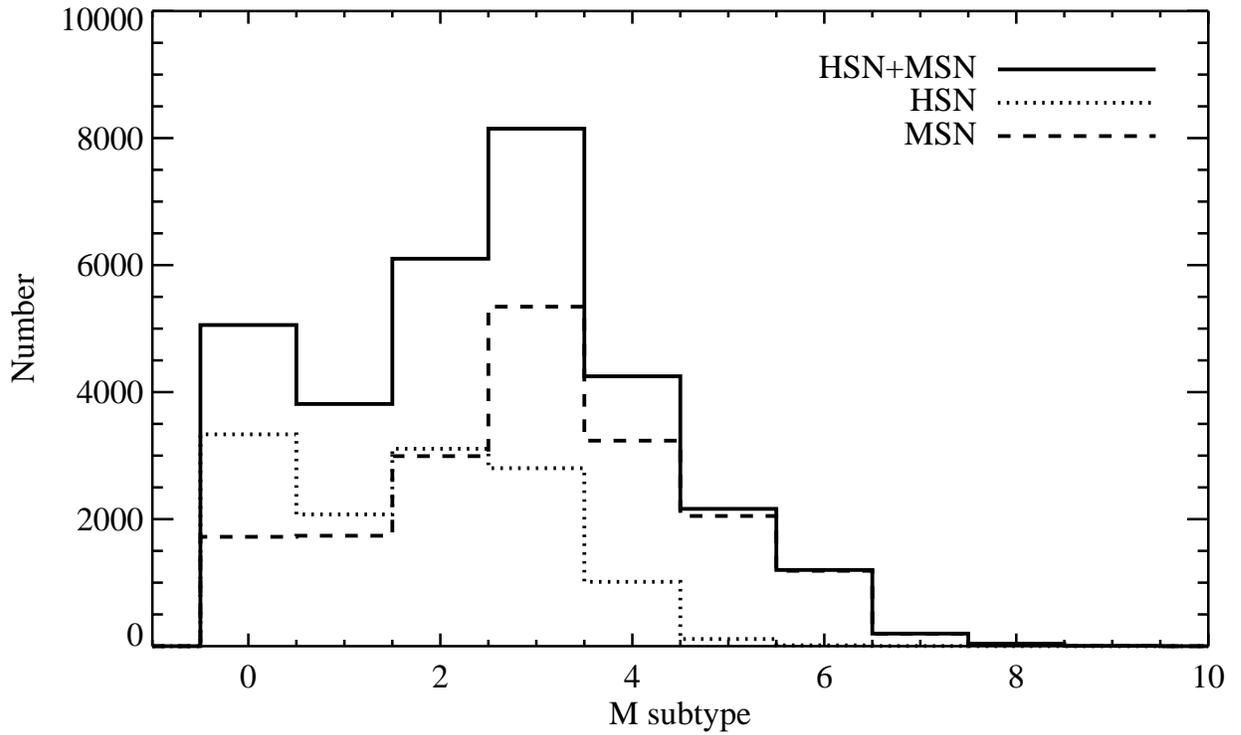}
  \caption{The distribution of spectral subtypes for the HSN (dotted), MSN (dashed), and total (solid) samples.
   As expected, the HSN sample mostly contains the earlier, and intrinsically brighter, spectral subtypes. 
  Although the MSN sample has more later spectral subtype stars, there are relatively few beyond M6.}
  \label{fig:spt_by_s2n}
\end{figure}

We made our cuts based on the continuum level in regions near the lines of 
interest because accurate emission line measurements 
depend on our ability to define the local continuum.
The high signal-to-noise (HSN) sample includes objects where all component spectra have median continuum levels at least three times 
the standard deviation of the points in the continuum regions for both H$\alpha$ and H$\beta$.
There are 43,788 spectra in the HSN sample, from 12,459 individual stars.
Because earlier-type stars are both bluer and brighter, the HSN objects 
are almost entirely spectral types M0-4.
The spectral type distribution of HSN spectra is shown in Figure \ref{fig:spt_by_s2n}.
 
Our medium signal-to-noise (MSN) sample is defined in a similar way 
to the HSN sample, except that we only require the median H$\alpha$ 
continuum level to be at least 3 times the standard deviation of the 
points in the region.
The MSN spectra have very noisy continuum flux near H$\beta$. 
However, many of these objects exhibited H$\beta$ line emission that can
be used in our flare analysis.
Like the HSN sample, all of the spectra for each object were required to 
meet the MSN criteria in order to be included in the MSN sample.
There are 68,203 spectra from 18,512 individual stars in the MSN sample. 
The spectral type distribution of the MSN spectra is also shown 
in Figure \ref{fig:spt_by_s2n}.

For completeness, we note that for $\sim$20\% of the stars in the 
original sample, the 
individual component spectra have low SNR that did not meet the MSN cut.
This LSN sample may well contain flares but cannot be used in the detailed 
analysis presented here.

\subsection{Spectral Type Determination}

The spectral type of each star was determined by \citet{West2008} from the co-added DR5 spectra using the Hammer 
spectral typing facility \citep{Covey2007}.
The Hammer uses measurements of molecular bands and line strengths to estimate a spectral type and is accurate to within one subtype.
We adopted the \citet{West2008} types for our analysis of the individual component spectra, which have lower SNR than the co-added spectra.

The Hammer is tuned to identify and classify M dwarfs during
quiescence by fitting templates to molecular band depths and overall
spectral shape.  Large flares produce optical continuum emission that
veils the molecular features and makes the spectral shape
significantly bluer.  Therefore, we verified that the sample does not
exclude continuum flares that cause the star to be misidentified by
the Hammer. We spectral typed the dM4.5e flare star YZ CMi during quiescence and during 
a giant flare of $\Delta$U $\sim$ -6 $mags$  \citep{Kowalski2010}. 
The quiescent spectral type returned by the Hammer was M4, while 
during some stages of the flare, types of M2 and M3 were found, still
reasonably close to the quiet value.
To further establish that the spectral type determination was 
not biased against flares, 
we also searched for flares in all spectroscopic objects whose
photometric colors (taken at a different time than the spectroscopy) were consistent with M dwarfs \citep{West2008,Kowalski2009},
but were not identified as M dwarfs by the Hammer.
There were no large, continuum-enhancement flares identified among 
these objects.
These two tests confirm that we did not select against continuum flares by relying on automatic spectral typing.

\section{Flare Analysis Tools}
\label{sec:flare_tools}

The unique SDSS data set provides an opportunity to observe the spectroscopic time evolution of a large number of flares.
It does, however, present some challenges.
First, the total length of consecutive observations is generally short ($\sim$45 minutes), and if a flare does occur, it may start before or at any time during the individual observations. 
Second, the time resolution is coarse: the duration of small flares is often less than the exposure time of one observation. 
Both of these factors make determining the time of the flare beginning, peak, or end ambiguous.
Additionally, only the largest flares, which are rare, will be evident throughout an entire sequence of spectra.
In the case of large flares that last for several hours, the observations may only cover a small fraction of the entire flare light curve.
Finally, flares are more likely to be observed during the longer gradual phase, 
when the continuum enhancement is likely to be significantly diminished.  
On the other hand, the emission lines will remain enhanced throughout the flare.  
Given these challenges, we chose to concentrate on identifying flares based on strong
and/or variable chromospheric line emission.

H$\alpha$ has been studied extensively in low mass stars, and is well known to vary on short timescales  \citep{Bopp1978,Gizis2002,Cincunegui2007, Walkowicz2009, Lee2010}, even outside of flares.
Thus, if H$\alpha$ is solely used to identify flares, there is no way of distinguishing variability from flaring (except for the largest H$\alpha$ flares), 
given the small number of exposures and the relatively long exposure 
times in our sample.
Although H$\beta$ variability has not been as well-studied on M dwarfs,
any flare large enough to be seen in a 15-minute exposure should also 
show an increase in H$\beta$ emission, as evidenced by studies of 
individual flares,
e.g. \citep{Mart'in2001,Fuhrmeister2004,Crespo-Chac'on2006, Kowalski2010}.
Therefore we required both the H$\alpha$ and H$\beta$ lines to meet our flare criteria for
line strength and/or variability (see \S \ref{sec:def_flare} for details of the flare criteria).
This requirement has the additional benefit of ensuring that cosmic rays are not falsely classified as flares.

\subsection{Flare Line Index}
\label{sec:FLI}

\begin{figure}
  \includegraphics[height=.5\textheight]{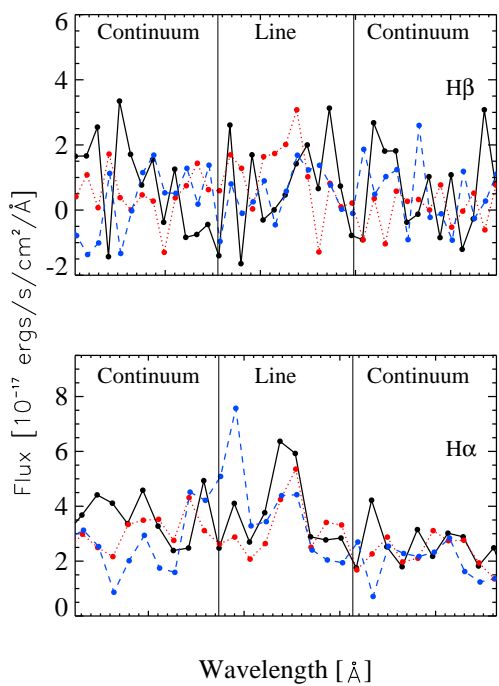}
  \caption{Three consecutive spectra (ordered by time: black/solid,
    red/dotted, blue/dashed) of an M5 star from the MSN sample that
    show no evidence of strong line flux in the H$\beta$ (top) and
    H$\alpha$ (bottom) line regions.
  The vertical lines mark the line regions used in the calculations.
  The EW values for the H$\beta$ are spuriously high in some exposures, while
  all of the FLI values are low. 
  For the H$\alpha$ line, the FLI values are similar to the EW measurements.
  This demonstrates the usefulness of using FLI values when the continuum is weak or uncertain.}
  \label{fig:false_ew}
\end{figure}

A commonly used measurement of emission line strength is the equivalent width (EW).
However, EW measurements are problematic when the continuum flux near the line is weak
and has low SNR.
Figure \ref{fig:false_ew} shows the H$\alpha$ and H$\beta$ line regions 
during a sequence of exposures of an M5 dwarf from our MSN sample. 
The H$\beta$ continuum flux is very close to zero, so division by the continuum 
to obtain EWs gives values of -3\AA, 106\AA, and 19\AA \hspace{1 pt} 
for the three consecutive exposures.
While the star in Figure \ref{fig:false_ew} is clearly not in a flaring state, the EW measurements for H$\beta$ are large and variable, 
as they might be during a flare.
This issue could be avoided by limiting the sample to objects with high SNR in the H$\beta$ continuum (i.e., the HSN sample).
However, including the MSN sample greatly increases our sample size.
We therefore developed a new method for measuring line strength that is independent of the continuum strength.
We define the Flare Line Index (FLI)  as:
\begin{equation}
\mbox{FLI} = \bar{l} / \sigma, 
\label{eqn:fli}
\end{equation}
where $\bar{l}$ is the mean value of the flux in the line region (for a specific emission line) minus the continuum flux (associated with the emission line; see Table 1) and $\sigma$ is the standard deviation of  the continuum. 
The FLI measures the line strength in terms of its significance compared to the noise.
 
The FLI allows us to determine whether a line measurement is statistically significant even when the continuum is weak and/or uncertain.
For the star in Figure \ref{fig:false_ew}, the FLI values for H$\beta$ are -0.3, 1.1, and 0.2, respectively.
The differences in these values can be attributed to the noise in the 
spectra.  The FLI values are neither as large nor as variable as the EW 
measurements of the same data. 
When the continuum is well-measured, as in the H$\alpha$ line, the FLI values and EW measurements are similar.
The FLI does depend on the SNR of the spectra; stars with higher SNR 
have larger FLI
values for the same emission line flux.

FLI values for both H$\alpha$ and H$\beta$ were computed for all spectra in the HSN and MSN
samples and used in the subsequent analysis.

\subsection{False Discovery Rate Analysis}
\label{sec:fdr}

Determining if an object is in a flaring state from a single spectrum
is difficult because active stars have a range of nominal,
non-flaring levels of H$\alpha$ activity. Also, there may be 
little difference between the 
emission line strengths of an active star in quiescence and those of 
a flaring star 
during a small flare or in the gradual phase of a large flare.  We 
searched for flares
based on the strength of the line emission, and on the variability in repeated 
spectral observations.  We did not require that both criteria be met, since
small flares
may not have particularly strong lines, while large flares may not show much
variability over the timescale of the SDSS observations.  

We used False Discovery Rate \citep[FDR;][]{Miller2001} analysis to search for spectra with 
particularly strong emission lines that were outliers from the typical stellar
distribution of FLI values.  Our sample exhibits a continuous distribution of measured FLI values for both H$\alpha$ and H$\beta$, 
ranging from lines in absorption through emission and into the flare regime.
FDR analysis is a quantitative method for determining a threshold value of a test statistic that separates real sources from a background (null) distribution for a given false-discovery rate.
In our case, we sought to determine the FLI value above which stars were considered to be in a flaring state.
The analysis uses an adaptive technique that establishes the threshold value based on
the desired false positive rate, defined by the parameter $\alpha$.
Our FDR method replaces an arbitrarily chosen confidence level (such as 3$\sigma$) with a
known false positive rate and can select the outliers from our sample 
(as compared to the null) as flares while controlling the rate of false 
detections through the choice of $\alpha$.
For our analysis, we limited the false positive rate to $\alpha = 10\%$.

The key to using FDR analysis is to assemble a meaningful null distribution. 
In the case of identifying flares, the null distribution should be a
set of spectra that have similar properties to our candidate flare
sample, but contain no flares. Active M dwarf spectra (that are not 
flaring) will provide a good null sample for the flare analysis.

\subsection{Constructing an Active Sample using FDR}
\label{sec:emission_sample}

The fraction of stars that are magnetically active increases from a few percent at 
spectral type M0
to nearly 100\% at type M9 \citep{Hawley1996,Gizis2000,West2004,West2008}.
Rather than simply adopting the active sample found by 
\citet{West2008}, we tested the effectiveness
of the FLI values and FDR analysis tools on our HSN and
MSN samples by verifying that we could reproduce the \citet{West2008}
results.  Again, we require a null distribution to separate
the active stars from the inactive ones.  For this case,
we adopted the DR5 composite spectra of inactive stars 
found by \citet{West2008} as the null distribution. We used 
the individual component spectra in the HSN and MSN samples
as the test distribution.  We carried out the analysis
separately for each spectral type, although we found that
the HSN sample only had sufficient numbers for the M0-3
types and the MSN sample for the M0-7 types
(see Figure \ref{fig:spt_by_s2n}).  

\begin{figure}
  \includegraphics[height=.5\textheight]{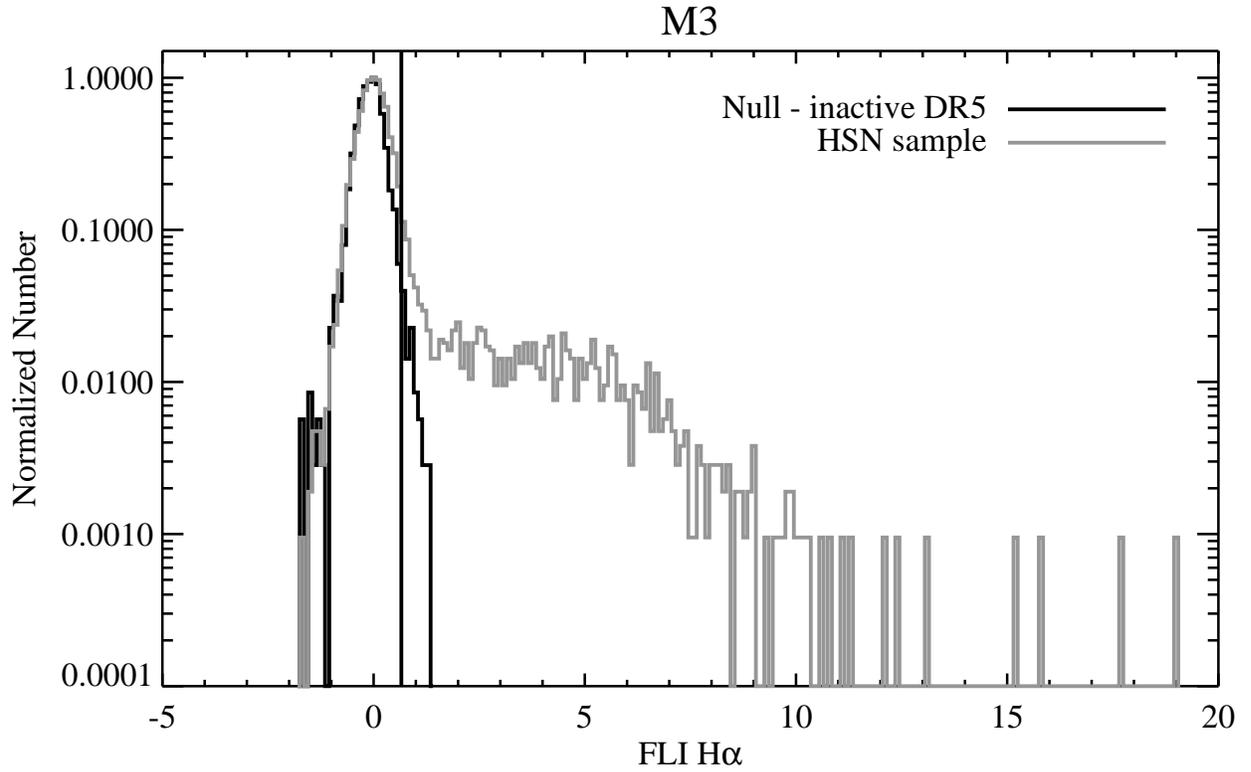}
  \caption{The distribution of FLI values for H$\alpha$ of the component spectra of M3 stars in the HSN sample (gray line) and the distribution of the null sample 
  taken from the inactive DR5 integrated spectra (black line). 
  The FDR-derived threshold value (vertical line) separates the inactive from the active and flaring spectra, 
  which are seen as the high FLI value tail of the HSN sample.
      }
  \label{fig:fdr_example}
\end{figure}

An example of the FDR analysis for M3 stars is shown in 
Figure \ref{fig:fdr_example}.
The gray line represents the distribution of FLI values for
the H$\alpha$ emission of the M3 dwarfs in the HSN sample (the test sample,
which includes both active and inactive stars),
while the black line represents the FLI distribution of the inactive
M3 dwarfs from the DR5 composite spectra (the null sample, which contains
only inactive stars).
The distributions have been normalized to have the same peak value
centered at FLI=0.
The vertical line (at a FLI of 0.66) is the threshold value determined
by the FDR analysis that divides active stars from inactive ones, with
the false positive rate $\alpha$ set to 10\%.  
The FLI emission threshold values we determined for the M0-3 spectral
types in the HSN sample are 
shown in Figure \ref{fig:fli_cutoffs} (labeled ``Emission''). 
Note that for later types, the M3 value was adopted.
Table 2 gives the H$\alpha$ FLI emission threshold values for 
both the HSN and MSN samples. 

Using our FLI/FDR determination of activity, we found that
2.8\% of the M0s are active and this fraction rises to 84\% of M7-9 stars.
These compare well to the active fractions reported by \citet{West2008}, who 
found 2.6\% and 90\%, respectively.
Although this agreement might be expected 
because we used the DR5 inactive spectra as our null distribution,
we used a different test statistic (FLI instead of EW) and
found our active sample using FDR analysis instead of defining an EW 
activity threshold of 1\AA \hspace{1 pt} on statistically significant emission lines 
in high SNR spectra \citep[see][for more details]{West2004}.
The fact that we successfully recovered the same active stars gives us 
confidence that FLI values are a 
useful measure in cases where the EW is ill-defined, and that the 
FDR method can be used to 
distinguish between the distributions. 

\begin{deluxetable}{lrrrrrrrrr}
\label{tab:fdr_thres}
\tablewidth{0pt}
\tablecaption{FLI threshold values} 
\tablehead{
\colhead{}    & \multicolumn{5}{c}{H$\alpha$}    & \colhead{} & \multicolumn{2}{r}{H$\beta$} & \colhead{} \\ 
\cline{2-6} \cline{8-10} \\ 
\colhead{}  &  \multicolumn{2}{c}{MSN} & \colhead{} & \multicolumn{2}{c}{HSN}  &  \colhead{} & \colhead{MSN} & \colhead{} & \colhead{HSN} \\
\cline{2-3} \cline{5-6} \cline{8-8} \cline{10-10} \\ 
\colhead{SpT} & \colhead{Emission} & \colhead{Flare} & \colhead{} & \colhead{Emission}  & \colhead{Flare} & \colhead{} &
	   \colhead{Flare}   &  \colhead{} & \colhead{Flare}}

\startdata
M0  &   1.12    &   2.97    &&    1.29    &    8.03    &&        0.61    & &       3.44         \\ 
M1  &   0.95    &    4.23    &&    1.09    &    10.12    &  &      1.50    &  &     4.74          \\ 
M2  &   0.96    &    7.76    &&   0.88   &   7.73    &  &      2.20    &  &     7.43       \\ 
M3   &   1.15    &    8.20    &&    0.66    &   12.13    & &       2.94    & &     10.70          \\ 
M4  &   0.76    &   9.73    &&   ...    &   ...    &  &   6.12        &&   ...         \\ 
M5  &   0.55    &   14.41    &&   ...    &   ...    &   &    8.10        &&   ...          \\ 
M6  &   0.68    &   10.80    &&   ...    &   ...    &   &    7.52        &&   ...          \\ 
M7  &   0.25    &   12.18    &&   ...    &   ...    &   &    5.45        &&   ...          \\ 
\enddata

\end{deluxetable}

\section{Flare Criteria}
\label{sec:def_flare}
We used two separate methods to define whether or not an object in our sample contained a flare.
The first was to use FDR analysis to determine which spectra had emission in both H$\alpha$ and H$\beta$ that was strong enough to be considered a flare.
The other method was to find objects whose H$\alpha$ and H$\beta$ lines were both 
in emission and showed large variations together with time.  These
two methods are not mutually exclusive; flares that have strong
emission lines that vary over the course of several exposures will 
meet both criteria.

\subsection{Strong Emission Lines using FDR Analysis}
\label{sec:fdr_flares}

We used the DR5 composite spectra of the active stars (i.e. those
that met the emission threshold criteria determined from \S \ref{sec:emission_sample})
as the null distribution for the flare FDR analysis.  

Although the objects that make up the null
distribution (the composite spectra) are the same objects that make up
the test distribution (the individual exposures), the distributions
are distinct in two significant ways.  First, the signature of small
flares that are only present in a subset of the exposures is diluted
in the composite spectra.  Second, outliers were removed from the null
distribution by rejecting spectra that were not part of the continuous
distribution of FLI values.  A few small or diluted flares may still
be present in the null distribution.  However, any flares in the null
will simply increase the threshold value determined in the analysis
and decrease the number of objects classified as flares.

\begin{figure}
  \includegraphics[height=.5\textheight]{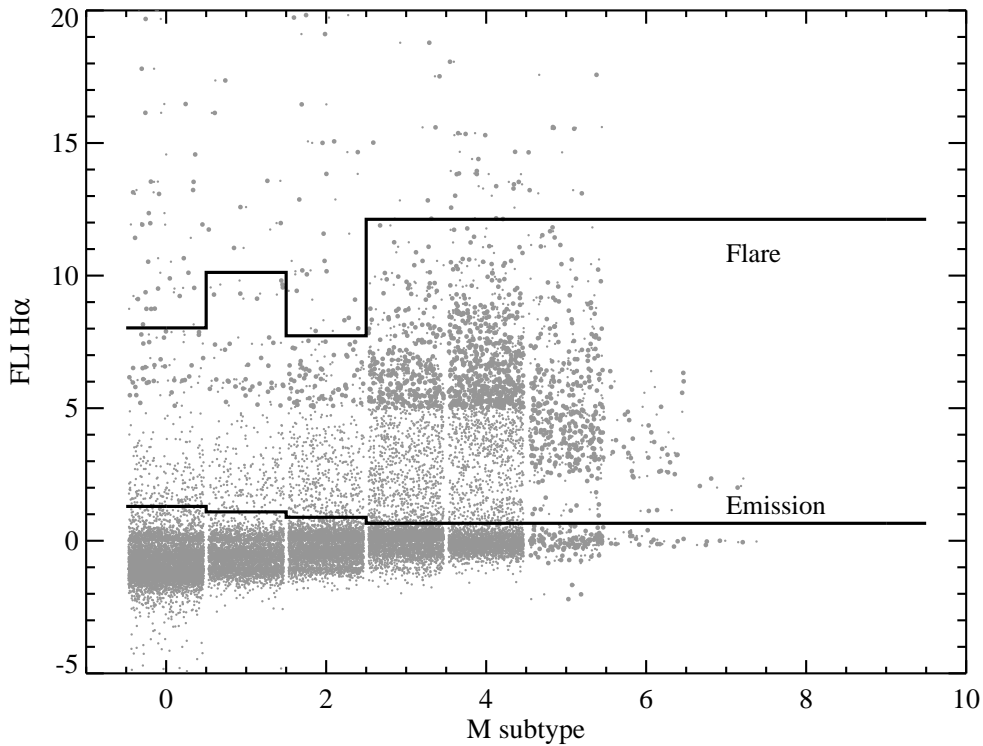}
  \caption{ The H$\alpha$ FLI values from the HSN sample as a function of spectral subtype are shown as gray points.
  Each point has been given a random offset within its spectral type bin and enlarged where the point density is low for clarity.
  Most spectra have low FLI values, with the mean increasing with later spectral type, as expected from the activity fraction studies of 
  \citet{West2004,West2006,West2008}.
  The solid line labelled ``Emission'' indicates the FDR-derived threshold values separating active spectra from inactive ones, 
  while the ``Flare'' line denotes the threshold values separating active stars from flares.
  Because our flare criteria require both H$\alpha$ and H$\beta$ to be above the relevant threshold value, not all points above
  the ``Flare'' line are categorized as flares in our sample.
  }
  \label{fig:fli_cutoffs}
\end{figure}

As in the active sample determination described in \S \ref{sec:emission_sample},
we first normalized the null and test distributions, and then
applied the FDR analysis separately to the HSN and MSN samples,
for each M subtype. Again we found that the HSN sample analysis was limited
to M0-3 types and the MSN sample to M0-7 types.  The flare
threshold values for both H$\alpha$ and H$\beta$ are given 
in Table 2, and are shown for the HSN sample in Figure \ref{fig:fli_cutoffs}.
To be classified as a flare, we required that an individual spectrum 
had FLI values for both H$\alpha$ and H$\beta$ above the respective
FDR flare threshold values.   Thus, many of the spectra above the Flare
line in Figure \ref{fig:fli_cutoffs} were not found to be flares because they did not
make the H$\beta$ cut.  This effectively removed cases of H$\alpha$
activity that were not strong enough to be classified as flares.

\subsection{Variable Emission Line Strength}
\label{sec:var_em_strength}
We also identified flares by looking for variation in emission line strength among individual exposures of the same object.
H$\alpha$ emission 
is known to vary outside of flaring (see discussion in \S \ref{sec:flare_tools});
\citet{Gizis2002} and \citet{Lee2010} 
found that ``quiescent'' H$\alpha$ emission often varies by as much as 30\%
on short timescales (minutes-hours).
This may well be due to low-level flaring, but it is not possible to classify
unequivocally as flare emission due to the relatively poor time resolution of
their data (and ours).  We therefore concentrated on variability that exceeds
this level (representing larger, definite flares), and developed the following
flare criteria based on both the H$\alpha$ and H$\beta$ FLI values: 

\begin{enumerate}
\item 
Mean FLI values $>$ 3 over all exposures for an object for both H$\alpha$ and H$\beta$; in these strong emission line spectra, we 
identified flares if both the H$\alpha$ and H$\beta$ FLI values for an
object differed by more than 30\% between the minimum and maximum values
measured over the course of the exposures.
\item
Mean FLI values $<$ 3 in both H$\alpha$ and H$\beta$; we identified flares in
these weak emission line data using the criterion that max FLI - min FLI $>$ 3. 
We adopted an absolute change criterion because a 30\% change is still 
within the noise when the mean FLI value is small.
Since FLI is defined as $\bar{l} / \sigma$,
this is essentially a 3$\sigma$ requirement (max FLI is 3$\sigma$ larger than
min FLI).
\item
Mean FLI values $>$ 3 in H$\alpha$ but $<$ 3 in H$\beta$;
these objects have moderate emission strength in H$\alpha$ but weak emission in H$\beta$.  We
require that they meet both
of the above variability criteria 
(max FLI - min FLI $>$ 3; percent change $>$ 30\%) to be considered a flare.
\end{enumerate}

\begin{figure}
  \includegraphics[width=.8\textheight]{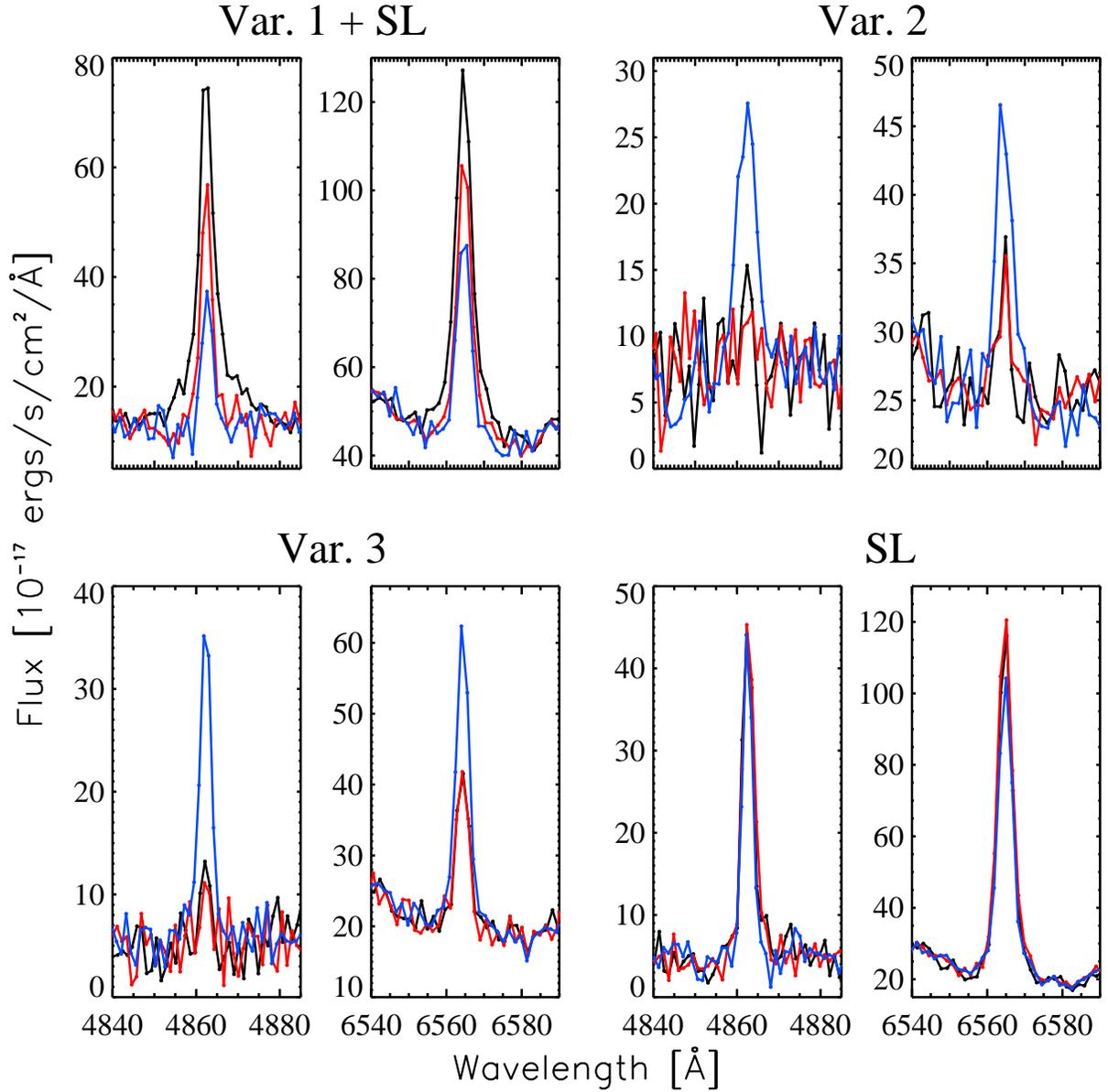}
  \caption{The H$\beta$ and H$\alpha$ lines of typical flares that meet each of the flare criteria 
  (see \S \ref{sec:fdr_flares} and \ref{sec:var_em_strength} for details about the strong lines and variability criteria).
   Each panel shows all of the exposures for that object, 
  with the black, red, and blue lines corresponding to the first, second, and third exposures, respectively. 
  Note that the flare that meets the first variability criterion also meets the FDR strong lines criterion. 
 }
  \label{fig:flare_crit_examples}
\end{figure}

We examined objects that narrowly missed passing these flare criteria
and confirmed that these quantitative measures are robust, and no real flares
were omitted from the sample.
Examples of flares that were identified based on each of these three conditions 
are shown in Figure \ref{fig:flare_crit_examples}. 
A large, decay-phase flare (top left) that met the first variability criterion (V1) also met the FDR strong lines criterion.
The example flare that met the second variability criterion (V2, top right) demonstrated weak emission during the first two exposures, but had significantly stronger 
emission in the final exposure. 
The moderate quiescent emission example (bottom left) met the third variability criterion (V3) with a flare during the third exposure.
The flare that met the FDR strong lines criterion (bottom right) showed very little change between exposures.
 
\section{Results}
\label{sec:results}

Our automatic flare identification using the above methods resulted in 285
individual spectra on 72 M dwarfs being classified as flaring.
We visually examined each spectrum and rejected 9 objects due to 
calibration errors, white dwarf companions, or cataclysmic variable signature.
Our final flare sample therefore comprises 243 spectra on 63 stars.
The properties of the 63 flaring stars are listed in Table 3.

\begin{deluxetable}{rrrrrll}
\label{tab:flares}
\tablewidth{0pt}
\tablecaption{Properties of the flares in the SDSS Spectroscopic Sample}
\tablehead{
\colhead{RA} & \colhead{Dec} & \colhead{SpT} & \colhead{Sample} & \colhead{\# exp.\tablenotemark{a}} & \colhead{ID Method\tablenotemark{b}} & \colhead{Notes\tablenotemark{c}} }

\startdata

 40.71900    &  1.05628    & M0   &  HSN    &  3    & SL     &     \\
355.34763    & -0.64664    & M1   &  HSN    &  5    & V1     &     \\
212.62507    & 38.67526    & M2   &  HSN    &  3    & SL     &     \\
  6.27294    &  0.26177    & M3   &  MSN    &  3    & SL     &     \\
 28.15489    &  0.94377    & M3   &  HSN    &  3    & SL + V1     & rise    \\
128.51469    & 23.71895    & M3   &  HSN    &  3    & V1     &     \\
134.89224    & 28.22401    & M3   &  HSN    &  3    & V1     &     \\
137.21883    & 25.20690    & M3   &  HSN    &  3    & SL + V1     & decay    \\
154.13328    & 36.06647    & M3   &  HSN    &  3    & V1     &     \\
161.54023    & 42.75901    & M3   &  HSN    &  6    & V1     & decay + quiet    \\
179.32923    & 36.89933    & M3   &  HSN    &  4    & V1     &     \\
231.35810    & 35.70936    & M3   &  HSN    &  3    & V1     &     \\
246.65711    & 34.37560    & M3   &  MSN    &  3    & V2     & has quiet    \\
  8.11627    & -0.66997    & M4   &  HSN    &  3    & V1     &     \\
 13.71803    &  0.52899    & M4   &  MSN    &  3    & V3     & has quiet    \\
123.54478    &  7.83326    & M4   &  HSN    &  4    & V1     &     \\
131.82185    &  3.01438    & M4   &  HSN    &  8    & V1     & has quiet    \\
138.47179    & 25.96210    & M4   &  HSN    &  3    & SL     &     \\
138.91181    &  9.04170    & M4   &  MSN    &  6    & V3     & has quiet    \\
151.16178    & 39.81706    & M4   &  HSN    &  3    & V1     & decay + quiet    \\
178.31625    & 35.11017    & M4   &  HSN    &  3    & V1     & rise    \\
232.12920    & 38.05164    & M4   &  MSN    &  4    & SL     &     \\
237.12605    & 51.50901    & M4   &  HSN    &  3    & V1     & has quiet    \\
238.82210    & 33.31065    & M4   &  HSN    &  3    & V1     &     \\
358.33475    &  0.27060    & M4   &  MSN    &  3    & SL + V1     & decay    \\
 24.02448    & -1.12992    & M5   &  MSN    &  5    & SL + V1     & has quiet    \\
114.81201    & 31.09497    & M5   &  MSN    &  3    & V3     & rise + quiet    \\
115.68026    & 35.39034    & M5   &  MSN    &  12    & V3     & has quiet    \\
116.05362    & 41.68687    & M5   &  MSN    &  8    & V3     & decay + quiet    \\
116.72870    & 36.04137    & M5   &  HSN    &  12    & V1     &     \\
123.96988    & 32.28493    & M5   &  MSN    &  3    & V3     & rise + quiet    \\
129.77425    &  6.91552    & M5   &  MSN    &  4    & V3     & peak    \\
136.04183    & 42.06829    & M5   &  MSN    &  4    & V1     &     \\
136.09474    &  0.26750    & M5   &  MSN    &  3    & SL     &     \\
156.27249    & 57.84446    & M5   &  MSN    &  3    & V2     & has quiet    \\
163.10690    & 37.36536    & M5   &  HSN    &  3    & V1     & rise    \\
181.06226    & 32.77287    & M5   &  HSN    &  3    & SL + V1     & rise + quiet    \\
196.07931    & 15.21429    & M5   &  MSN    &  3    & V1     & decay    \\
208.73389    & 48.89529    & M5   &  MSN    &  4    & V1     &     \\
211.05023    &  4.04835    & M5   &  MSN    &  3    & V3     & peak    \\
213.01983    & 54.68985    & M5   &  MSN    &  3    & V1     & has quiet    \\
217.63041    & 57.17282    & M5   &  HSN    &  3    & V1     &     \\
219.60447    & 39.63877    & M5   &  MSN    &  4    & V3     & rise + quiet    \\
225.05331    & 60.90313    & M5   &  MSN    &  3    & V1     & peak    \\
248.08604    & 23.94904    & M5   &  HSN    &  3    & V1     & rise    \\
  3.28885    & -0.43108    & M6   &  MSN    &  3    & SL + V1     &     \\
 53.72138    & -7.31779    & M6   &  MSN    &  3    & SL + V1     & cont. emission    \\
118.05603    & 27.77888    & M6   &  MSN    &  4    & V2     & rise + quiet    \\
130.79823    & 35.31980    & M6   &  MSN    &  3    & V1     & decay    \\
143.62620    &  3.87247    & M6   &  MSN    &  6    & V3     & decay    \\
148.58678    & 34.44622    & M6   &  MSN    &  3    & V1     &     \\
194.47690    &  3.55015    & M6   &  MSN    &  3    & SL + V1     & cont. emission    \\
195.65088    &  6.03001    & M6   &  MSN    &  4    & SL + V1     & decay    \\
240.34528    & 51.89663    & M6   &  MSN    &  3    & V1     & peak    \\
240.91342    & 38.51986    & M6   &  MSN    &  3    & SL     &     \\
329.51701    & -8.35545    & M6   &  MSN    &  3    & V3     & has quiet    \\
119.29499    & 42.94881    & M7   &  MSN    &  4    & SL     &     \\
135.52879    &  0.55538    & M7   &  MSN    &  4    & V1     & peak    \\
180.38979    & 40.77934    & M7   &  MSN    &  4    & SL + V1     & rise    \\
193.30205    & 40.56684    & M7   &  MSN    &  3    & SL     & decay    \\
234.07976    & 33.08753    & M7   &  MSN    &  3    & SL + V1     & decay    \\
 30.09824    &  0.64641    & M8   &  MSN    &  3    & SL     &     \\
182.07041    &  8.75788    & M9   &  MSN    &  4    & SL + V3     & has quiet    \\

\enddata
\tablenotetext{a}{This is the number with no CRs in either H$\alpha$ or H$\beta$ line regions}
\tablenotetext{b}{The method used to identify flares. SL is the strong line criterion (see \S \ref{sec:fdr_flares}) 
and V1,V2, and V3 are the three variability criteria (see \S \ref{sec:var_em_strength})} 
\tablenotetext{c}{``Rise'',``peak'', and ``decay'' phase flares are defined in \S \ref{sec:phases}. 
We also note flares in which at least one spectrum is in quiescence, as well as the two flares with enhanced continuum emission
 (discussed in \S \ref{sec:cont_enhancement}).}
\end{deluxetable}

 A summary of the number of stars and exposures that met each flare criterion is presented in Table 4, which also shows
that the majority of flares come from the MSN sample.  This is not unexpected,
since there are many more stars in that sample.
Our results suggest that flares in this sparsely sampled 
spectroscopic data set are most easily identified through 
emission line variability rather than the absolute strength of
the emission lines.

 \begin{deluxetable}{crrrrrrrrrrrrrrrrr}
\tabletypesize{\scriptsize}
\label{tab:method_id}
\tablewidth{0pt}
\tablecaption{Flare identification by emission line strength and variability 
criteria}
\tablehead{
\colhead{}    & \multicolumn{2}{c}{SL}    & \colhead{} & \multicolumn{2}{c}{V1} & \colhead{} & \multicolumn{2}{c}{V2} & \colhead{}
 & \multicolumn{2}{c}{V3} & \colhead{} & \multicolumn{2}{c}{SL + V1} & \colhead{} & \multicolumn{2}{c}{SL + V3} \\ 
\cline{2-3} \cline{5-6} \cline{8-9}  \cline{11-12} \cline{14-15} \cline{17-18} \\ 
\colhead{} & \colhead{Stars} & \colhead{Exp.} & \colhead{}& \colhead{Stars} & \colhead{Exp.}  & \colhead{} & \colhead{Stars} & \colhead{Exp.} &
	  \colhead{} & \colhead{Stars}   &  \colhead{Exp.} & \colhead{} & \colhead{Stars} & \colhead{Exp.} & \colhead{} & \colhead{Stars} & \colhead{Exp.}} 

\startdata
HSN  &   3 & 9 &&      18   & 75   && 0   & 0   && 0   & 0           && 3   & 9   && 0   & 0   \\
MSN  &   7 & 23 &&       9   & 30   && 3   & 10   && 11  & 55     && 8   & 28   && 1   & 4   \\
Total  &  10 & 32 &&     27 & 105 && 3 & 10 && 11 & 55           && 11 & 37 && 1 & 4 \\

\enddata

\end{deluxetable}

\subsection{Time Evolution of Flares}
\label{sec:verification}

We first analyzed our flare sample to verify that the properties of the 
flares we identified are consistent with the characteristic time evolution 
seen in continuous monitoring observations.  Because our data might capture 
any or all phases of a flare, we 
expected to see examples of flares in the rise, peak, and decay phases
as well as a few large impulsive phase flares that show 
blue continuum flux enhancement. 

\subsubsection{Rise, Peak, and Decay Phase Flares}
\label{sec:phases}
We defined decay-phase flares as those showing both H$\alpha$ and H$\beta$ in 
at least 
two consecutive exposures with diminishing FLI values. 
Rise-phase flares were defined in the opposite sense, with two or more 
consecutive exposures having increasing FLI values for both
H$\alpha$ and H$\beta$.
There were ten rise phase flares and 11 decay phase flares in the sample.
Additionally, we saw five ``peak'' flares that show first increasing and then
decreasing FLI values, which we interpret 
as flares that were seen through most of their evolution.
We also identified flares that occurred in only a subset of the exposures,
either in the final exposure of a sequence, or in cases where some of the exposures are separated by many hours or days from the other exposures. 
In 19 cases, at least one exposure was obtained that showed a quiescent spectrum with no sign of flaring.
Flares with quiescent spectrum, as well as rise, decay, and peak phase flares are noted in the final column of Table 3.
The time evolution of the flares we have identified is consistent with
the known characteristics of flares.

\subsubsection{The Balmer decrements of Rise and Decay Flares}
Previous observations using time resolved spectroscopic monitoring 
have shown that the higher order Balmer lines show larger increases in flux 
during flares than the lower order lines.  This results in the observed Balmer 
decrement
(ratio of individual lines to a fiducial, often H$\beta$ as we adopt here)
becoming flatter, since H$\gamma$ and the higher order lines
strengthen relative to H$\beta$ 
\citep[][see also Figure 7 of Bochanski et al. 2007]{Garc'ia-Alvarez2002,Allred2006}
This behavior can be explained by a model where flare heating
increases the local electron density at chromospheric temperatures.
The higher order Balmer lines are less optically thick than the lower order
lines, so the increase in electron density causes a relatively larger increase
in their emission.
The H$\alpha$ line behavior relative to the other lines has previously 
been more difficult to measure during flares, since
observations typically concentrated on the blue part of the spectrum
\citep[e.g.][]{Hawley1991}. 
Along with their blue spectra, \citep{Eason1992} obtained data with a different instrument
at lower time resolution showing that the H$\alpha$ line,
being already very optically thick,
does not respond strongly during a large flare in accord with models
\citep{Allred2006}.

\begin{figure}
  \includegraphics[height=.5\textheight]{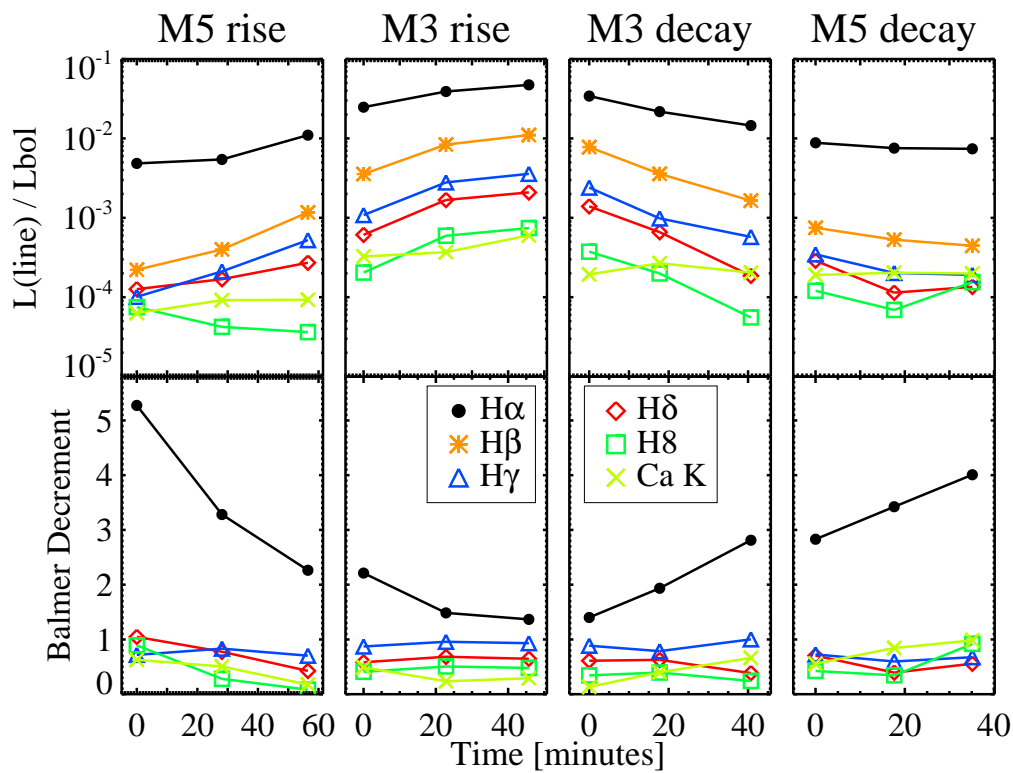}
  \caption{Line fluxes of emission lines normalized by the bolometric
    luminosity for two rise phase flares and two decay phase flares
    (top).  Note that Ca II K (crosses) does not follow the same trend as the Balmer lines.
  The bottom panels show the Balmer decrements for the same flares.
 The H$\alpha$ decrement is smaller when the flare is closer to its peak (later for rise phase flares, and earlier for decay phase flares).
   }
  \label{fig:balmer_dec}
\end{figure}

Figure \ref{fig:balmer_dec} illustrates the line flux normalized to the
bolometric luminosity for several lines (top), as well as Balmer decrements for four flares (bottom; two rise- and two decay-phase).
The Balmer lines increase during the rise phase flares, and decrease during the decay phase flares, as expected.
In the bottom panels, we see that for all of the Balmer lines except H$\alpha$, the decrements are flat throughout the flares.
As expected, the H$\alpha$ decrement decreases during the rise-phase, when
H$\beta$ is becoming relatively stronger, and increases during the decay-phase,
as the atmosphere gradually returns to its quiescent temperature and density
structure.  Our observations are in accord with expectations, and actually
represent truly simultaneous measurements with the same instrument
of the entire Balmer decrement (including H$\alpha$) during stellar flares.

The Ca II K line responds less strongly, and rises and decays more slowly 
during flares, than the Balmer lines
\citep{Butler1991,Hawley1991}, as is also evident from our observations.

\subsubsection{Continuum Flux Enhancement}
\label{sec:cont_enhancement}
\begin{figure}
  \includegraphics[width=.7\textheight]{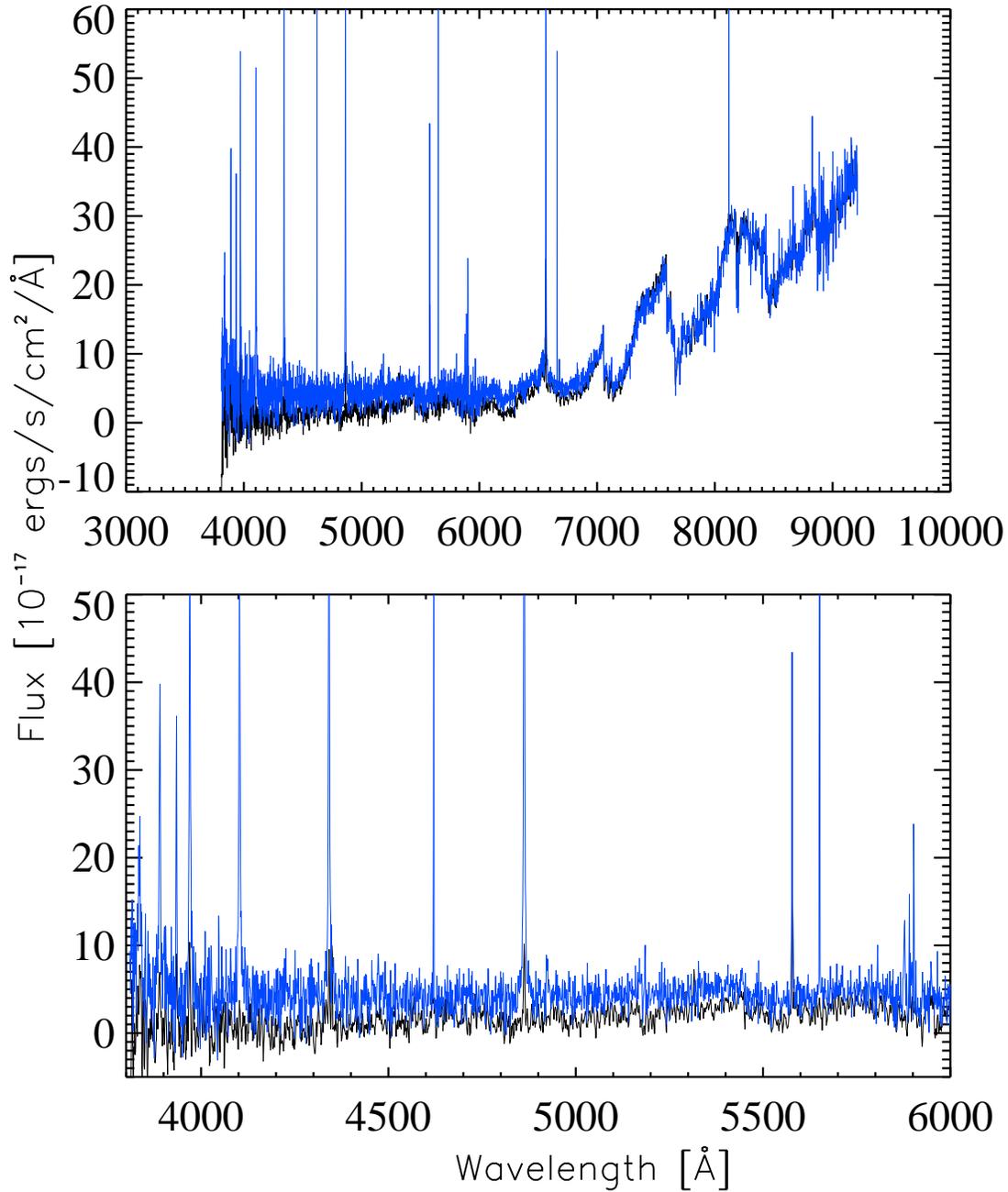}
  \caption{The spectrum of an M6 dwarf (R.A., Dec. = 194.47690, 3.55015) during quiescence (black) and during a flare (blue).
  This exposure captures at least part of the impulsive phase of the flare and shows a clear continuum enhancement in the blue portion of the spectrum (shown in the bottom panel),
   as well as strong emission lines.
  }
  \label{fig:flare_flux_523}
\end{figure}

Large flares cause strong blue continuum flux enhancement, with less 
dramatic effects at redder wavelengths.
Only very strong flares caught in the impulsive phase would lead to a
significant flux enhancement for the wavelength range and relatively long
exposure times (which dilute the short-lived continuum burst) of the SDSS spectra.
We expected to find only a small number of such flares, since large flares are rare and 
we are only sensitive to measuring continuum enhancement on HSN stars or very bright MSN stars, where we have a good measure of the quiescent continuum.
Our sample contained two continuum flares, an example of which is shown in Figure \ref{fig:flare_flux_523}.
Both flares occurred during the third and final exposure of a series, and both show very strong emission lines.
These are evidently large flares that were caught during the impulsive phase.
We measured the flare-only spectrum by subtracting the quiescent (first) 
exposure obtained for each star from the flare exposure.
We found that the continuum flux represented 88\% and 90\% of the total
flare radiation, respectively, with the remainder of the energy coming from
the line radiation.
\citet{Hawley1991} found that the continuum emitted 96\% of the flare 
flux during the impulsive phase of their very large flare, and averaged 91\% over the course of the flare.
This suggests that our flare exposures may have included the impulsive phase and 
part of the decay phase of each flare.
The existence of these continuum flare spectra in our sample, and the relative strength of the lines and continua, 
 give us additional confidence that we are correctly identifying flares
by our automatic algorithms.

\subsection{Flare Duty Cycle}

\begin{figure}
  \includegraphics[height=.5\textheight]{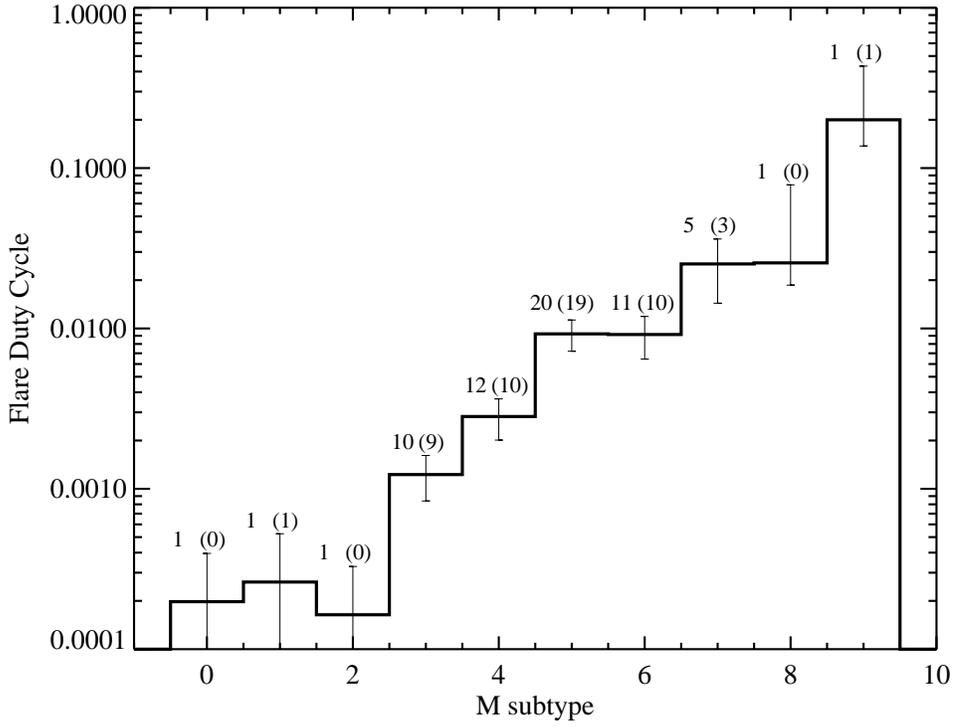}
  \caption{The flare duty cycle as a function of spectral type. 
  The error bars are calculated from the binomial distribution.
   The number of flares in each bin is given, with the number of flares that meet the variability criteria given in parentheses.
   The flare fraction increases with spectral subtype as expected from previous studies, and is similar to the rise in the active fraction of M dwarfs.
      }
  \label{fig:flares_spt}
\end{figure}

Figure \ref{fig:flares_spt} shows our observed flare duty cycle as a function of spectral type.
The duty cycle increases with later spectral type, from 0.02\% for M0 stars to 20\% for M9 stars. 
The error bars were calculated from binomial statistics, and the number of flares identified at each spectral type is noted.
The numbers in parentheses are the 
number of flares identified by the variability criteria. 
The much larger duty cycles for M7 and later are based on a small number of stars, and although there are many thousands of objects in the M0-M3 bins,
there are very few flares.
In order to improve the statistics, we binned by spectral type and found the duty cycles to be 
$0.022\pm0.016\%$ of M0-1 stars;  $0.077\pm0.023\%$ for M2-3 stars; $0.56\pm0.09\%$ for M4-6 stars; and  $3\pm1\%$ of M7-9 stars.

\begin{figure}
  \includegraphics[height=.5\textheight]{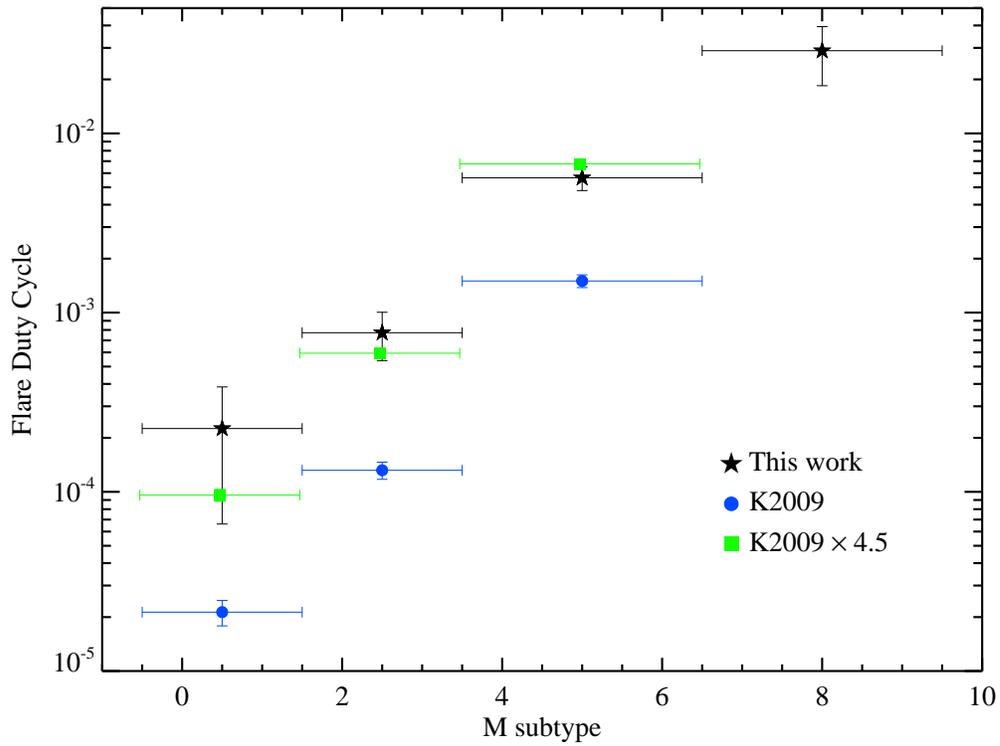}
  \caption{The flare duty cycle as measured by \citet[][closed circles]{Kowalski2009}
     and this study (stars), for spectral type bins M0-1, M2-3, M4-6, M7-9. 
  Scaling the \citet{Kowalski2009} results by a factor of 4.5 (squares) makes them consistent with our results.  
     }
  \label{fig:ejh_afk_rates}
\end{figure}

\citet{Kowalski2009} investigated flares in a large SDSS photometric sample.
They required near simultaneous photometric enhancements in both the $u$ 
and $g$ filters, and employed a threshold of 
$|\Delta u| >$ 0.7 to define a flare.
Figure \ref{fig:ejh_afk_rates} compares their flare duty cycle 
with our results for the binned data.
We find a higher duty cycle, by a factor of $\sim$ 4.5,
but the duty cycle as a function of spectral type has a similar trend in 
both studies.
It is not surprising that the duty cycles are different, since the two
studies use such different 
observations and threshold criteria. However, we
can qualitatively estimate the effect of observing line versus
continuum enhancements, and the effect of exposure time.  
The line emission remains enhanced for much longer than the continuum
emission leading to a higher probability of observing flares in line emission.
Longer exposure times increase the chance of a flare occurring during the
exposure.  Both of these factors lead to the observed higher duty cycle
seen in our spectroscopic observations.

\subsection{The Vertical Distribution of Flaring Stars in the Galaxy}

Nearly all flares occur on magnetically active stars, as defined by H$\alpha$ in emission \citep{ Reid1999, Kowalski2009}.
As discussed in Section \ref{sec:emission_sample}, the active fraction increases with spectral subtype.
In addition, \citet{West2008} found that the active fraction decreases with vertical height from the Galactic Plane
(denoted by $|$Z$|$) for all spectral types, which they attribute to an age
effect, where older stars are less likely to be active.  The age at which activity disappears
 changes with spectral type however, with later type M dwarfs remaining
active for a longer time.

We limited our analysis of the vertical distribution of the flaring stars
to just the M4-6 dwarfs, which show a large number of flares, and 
have a similar flaring fraction (see Figure \ref{fig:flares_spt}).
Earlier type stars have very few flares, and later type stars are too faint
to be observed at large distances.

\begin{figure}
  \includegraphics[height=.5\textheight]{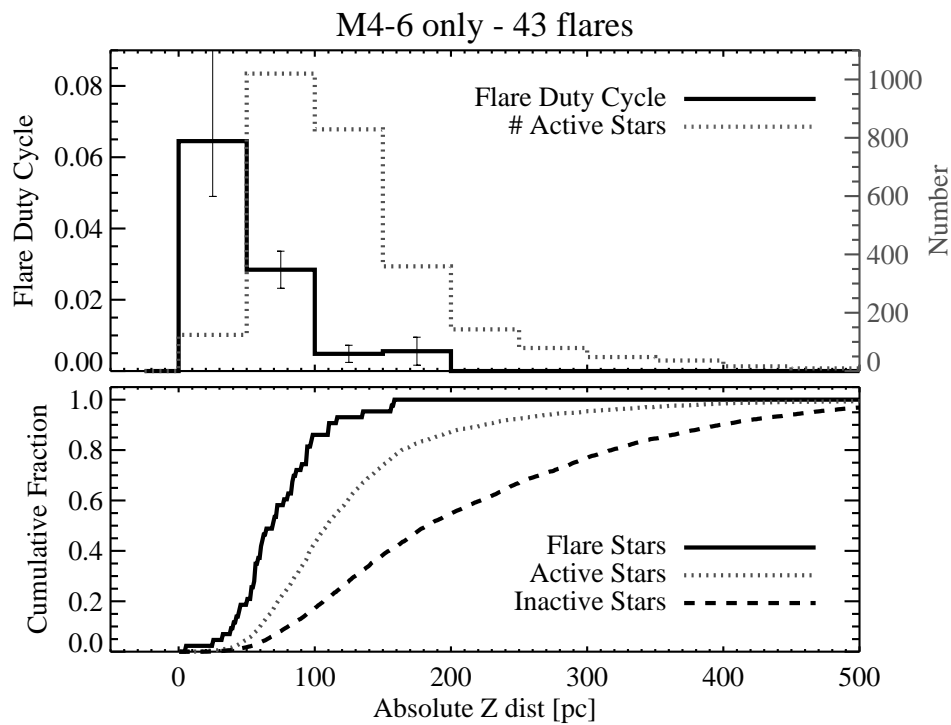}
  \caption{ Top: The flare duty cycle as a function of distance from the Galactic midplane ($|$Z$|$) for M4-6 stars (black line).
 The number of active stars \citep{West2008} in the sample is also shown (gray line, dotted, right axis).
    The fraction of objects that flare decreases rapidly with
    distance from the Galactic Plane.  Bottom: the cumulative distribution of absolute Z distance for the flare, active, and inactive stars between subtype M4 and M6.
    These results indicate that flaring occurs preferentially closer
  to the Galactic Plane than the general active star population, and therefore on younger stars. 
   }
  \label{fig:zdist}
\end{figure}

Figure \ref{fig:zdist} (top panel) shows the distribution of active M4-6 dwarfs 
and the fraction that showed flares as a function of vertical distance from the 
Galactic Plane.
We repeated our analysis, adjusting both the bin size and bin positions 
and verified that the increase in flare fraction as a function of $|$Z$|$
 is a real effect and not an artifact of the binning process.
The bottom panel of Figure \ref{fig:zdist} shows the cumulative distributions 
of the flare stars (black line), the active stars (dotted gray line) and 
the inactive stars (dashed line).
Although there are thousands of objects beyond $|$Z$|$=200 pc, none of them showed flares.
\citet{Kowalski2009} and \citet{Welsh2007} also found that all of the flares on 
M4 dwarfs occur within 200pc of the Plane.

Our results indicate that flares occur on stars that are preferentially closer 
to the Plane than the inactive stars, and closer even than the active
population.  If we interpret this in terms of an age
effect, it suggests that flare stars (which are a subset of the active 
population) are younger than
the magnetically active stars as a whole.  Evidently the frequency of
flaring decreases as
a star ages, although it still may exhibit quiescent chromospheric emission
and therefore be counted as active.

\section{Conclusion}
\label{sec:conclusion}

We have presented the first study to systematically search for flares in 
a sample of tens of thousands of time-resolved spectroscopic observations
from the SDSS Data Release 6.
We developed two new techniques for identifying and analyzing flares in
relatively low SNR spectra: the Flare Line Index (FLI) measurement 
and the application of 
the False Discovery Rate to flare spectra.
FLI values provide a measure of emission line strength even when the continuum level is not well-measured.
The FDR analysis distinguishes between the overlapping distributions in line strength of strong quiescent emission and genuine flares, providing 
a threshold FLI value for being considered a flare, while controlling 
the rate of false-positives.
Using our measured FLI values for the
H$\alpha$ and H$\beta$ lines, we applied the FDR analysis and a set of
variability criteria to find flares.

We identified flares on 63 stars from the 30,971 total stars comprising both the HSN and MSN samples.
The flares include examples of rise and decay phases,
as well as some sequences of exposures that appear to trace the flare through its entire evolution.
The Balmer decrements are
consistent with results from time resolved spectroscopic monitoring of individual flares and expectations of models.
We also found two flares that show continuum flux enhancements, which occur 
during the initial impulsive phase of large flares.
Because the properties of the flares in our sample are consistent with
previous findings, we are confident that our techniques, 
which we developed specifically to work with coarsely time-resolved spectroscopic observations, successfully identified flares.

We found that the flare duty cycle increases monotonically as a function of M dwarf spectral subtype, 
from $0.022\pm0.016\%$ of M0-1 stars to $3\pm1\%$ of M7-9 stars.
The flare distribution as a function of spectral subtype is similar to the distribution of active stars, providing further evidence that most flares occur on active stars.  
Nearly all of the flares in the sample occur on stars near the Galactic Plane, a result that is consistent with other studies
 \citep[i.e.][]{Kowalski2009}.
Furthermore, we have shown that flares occur preferentially on stars 
closer to the Plane than the mean active population, indicating that 
flare frequency likely decreases with age.

This study represents an important step toward understanding the
intrinsic flare frequency distribution as a function of stellar active
lifetimes. 
Surveys that don't have high time resolution and continuous coverage
(from which the total flare energy can be computed) require detailed and nuanced Monte Carlo simulations in order to constrain the 
physical flare frequency distribution.
Simulations which consider stellar samples in the Galactic context
is the subject of ongoing study (Hilton et al., in prep).

 The authors would like to thank J. Wisniewski, S. Schmidt, and J. Davenport for fruitful discussions that improved this paper.
 They would also like to thank D. Schlegel for assistance in obtaining the data.
 E.J.H., S.L.H. and A.F.K acknowledge support from NSF grant AST 08-07205.
 
 Funding for the SDSS and SDSS-II has been provided by
the Alfred P. Sloan Foundation, the Participating Institutions,
the National Science Foundation, the U.S. Department of
Energy, the National Aeronautics and Space Administration,
the Japanese Monbukagakusho, the Max Planck Society, and
the Higher Education Funding Council for England. The SDSS
Web site is http://www.sdss.org/.
The SDSS is managed by the Astrophysical Research Consortium
for the Participating Institutions. The Participating Institutions
are the American Museum of Natural History, Astrophysical
Institute Potsdam, University of Basel, University
of Cambridge, Case Western Reserve University, University 
of Chicago, Drexel University, Fermilab, the Institute for Advanced
Study, the Japan Participation Group, Johns Hopkins
University, the Joint Institute for Nuclear Astrophysics, the
Kavli Institute for Particle Astrophysics and Cosmology, the
Korean Scientist Group, the Chinese Academy of Sciences
(LAMOST), Los Alamos National Laboratory, theMax-Planck-
Institute for Astronomy (MPIA), the Max-Planck-Institute for
Astrophysics (MPA), New Mexico State University, Ohio State
University, University of Pittsburgh, University of Portsmouth,
Princeton University, the United States Naval Observatory, and
the University of Washington.

\clearpage

\clearpage

\clearpage

\clearpage

\clearpage

\clearpage

\clearpage

\clearpage

\clearpage

\clearpage

\clearpage

\clearpage

\end{document}